\documentclass[preprint2,twocolumn,times,tighten]{aastex63}

\usepackage{graphicx,times}
\usepackage{subfigure}
\newcommand{\be}{\begin{equation}}
\usepackage{threeparttable}
\usepackage{booktabs}
\newcommand{\ee}{\end{equation}}
\newcommand{\bea}{\begin{eqnarray}}
\newcommand{\eea}{\end{eqnarray}}

\usepackage{float}
\usepackage{amsmath}
\usepackage{cases}
\usepackage{longtable}
\usepackage{hyperref}
\usepackage{epstopdf}
\usepackage{amsmath,bm}
\usepackage{amssymb}
\usepackage{natbib}
\usepackage{morefloats}
\usepackage{multirow}
\usepackage{array}
\usepackage{verbatim}
\usepackage{color}
\usepackage{CJK}
\usepackage{lineno}
\begin{document}
\begin{CJK*}{UTF8}{gbsn}

\title{Toward interpreting the {\it IBEX} ribbon with mirror diffusion in interstellar turbulent magnetic fields}

\author[0000-0002-0458-7828]{Siyao Xu (徐思遥)}
\affiliation{Department of Physics, University of Florida, 2001 Museum Rd., Gainesville, FL 32611, USA; xusiyao@ufl.edu
\footnote{NASA Hubble Fellow}}

\author[0000-0003-3556-6568]{Hui Li (李晖)}
\affiliation{Los Alamos National Laboratory, NM 87545, USA; hli@lanl.gov}

\begin{abstract}

We investigate the role of the magnetohydrodynamic
(MHD) turbulence measured by Voyager in the very local interstellar medium (VLISM) in modeling the Interstellar Boundary Explorer ({\it IBEX}) ribbon. 
We demonstrate that the mirroring by compressible modes of MHD turbulence dominates over that by the mean magnetic field. 
Based on the new mirror diffusion mechanism identified by 
\citet{LX21} 
for particles with large pitch angles in MHD turbulence, 
we find that 
the mirror diffusion 
%{resulting from the perpendicular superdiffusion of turbulent magnetic fields} 
{can both confine pickup ions and preserve their initial pitch angles, 
and thus accounts for the 
enhanced intensity of 
energetic neutral atoms
that return to the heliosphere.}
The ribbon width is determined by both the range of pitch angles for effective turbulent mirroring and the field line wandering induced by Alfv\'{e}nic modes. 
{It in turn provides a constraint on the amplitude of magnetic fluctuations of fast modes.} 
The field line wandering also affects the coherence of the ribbon structure across the sky. 
By extrapolating the magnetic energy spectrum measured by 
{\it Voyager}, we find that the injection scale of the turbulence in the VLISM is less than $\sim 500$ au for the ribbon structure to be coherent.

\end{abstract}

\section{Introduction}

The ``ribbon" of enhanced energetic neutral atom (ENA) emissions discovered by 
the Interstellar Boundary Explorer ({\it IBEX}) 
\citep{Mcc09}
provides valuable information about the 
local interstellar magnetic field 
\citep{Schwad09,PogH11}
and the interaction of pickup ions with the pre-existing turbulent magnetic fluctuations outside the heliopause
\citep{Gia15,Zir20}.
Turbulent magnetic fluctuations following a Kolmogorov spectrum are observed by {\it Voyager 1} and {\it 2}
in the outer heliosheath 
\citep{Burla18,Zhao20,Bur22,Lee20,Frat21}. 
With a higher amplitude than that of the interstellar turbulent spectrum 
\citep{Burla18,Lee20,Ock21} and 
ion-neutral collisional damping of the interstellar turbulence in the partially ionized very local interstellar medium (VLISM)
\citep{XLi22}, 
the turbulence detected by {\it Voyager} is more likely to be of heliospheric origin
{\citep{Zank19}.} 
Among the numerous models for explaining the origin of the {\it IBEX} ribbon, 
the magnetic mirroring of pickup ions in the turbulent magnetic fields 
in the VLISM
is found to be important in determining the ribbon structure 
\citep{Gia15,Zir20}.

The magnetic mirrors created by 
summing over a large number of randomly
directed plane waves with 
random polarizations and phases
\citep{Giacalone_Jok1999}
can trap particles as particles are reflected back and forth between two mirror points
\citep{CesK73}.
Unlike compressible magnetohydrodynamic
(MHD) waves, MHD turbulence contains 
fast and slow (or pseudo-Alfv\'{e}nic) modes 
that generate magnetic mirrors and 
Alfv\'{e}nic modes that cause perpendicular 
superdiffusion of magnetic field lines
{\citep{CL03,Beresd13}}.
Following turbulent magnetic fields lines, particles also undergo superdiffusion in the direction perpendicular to the magnetic field 
\citep{XY13,LY14,Hu22cr}
while interacting stochastically with different magnetic mirrors along the magnetic field. 
As a result, {rather than being trapped,}
particles exhibit 
the {\it mirror diffusion} 
\citep{LX21}
parallel to the magnetic field.
The mirror diffusion is usually a much slower diffusion process compared to that associated with pitch-angle scattering 
\citep{LX21,Xu21}.

The MHD turbulence measured by {\it Voyager} contains both incompressible and compressible modes 
\citep{Zank19,Lee20,Frat21}.
In this work, we will 
investigate the importance of 
{the recently identified mirror diffusion of particles} in 
the MHD turbulence in the VLISM
in explaining the origin of the {\it IBEX} ribbon
and affecting its structure. 
In Section 2, we first examine the mirroring effect of the compressed interstellar mean magnetic field. In Section 3, we move to the mirroring effect of turbulent magnetic fields. 
The mirror diffusion {resulting from the perpendicular superdiffusion of turbulent magnetic fields}
and its effect on the ribbon width is discussed in Section 4. 
The effect of field line wandering on ribbon width and structure is studied in Section 5. 
Further discussion is presented in Section 6. A summary of our main results follows in Section 7.

\section{Mirroring effect of the interstellar mean magnetic field}
\label{sec: mirrmean}

The compression of the large-scale mean field in the outer heliosheath gives rise to the magnetic mirroring effect. 
It can cause the concentration of pickup protons in a narrow region where 
the radial component of the compressed interstellar mean magnetic field is close to zero
\citep{Mcc09,Chal10}.
%leading to a narrow region of enhanced pickup-proton intensity.
%accumulation/concentration 

For a particle with $r_g < L_\text{m}$, where $r_g$ is the particle gyroradius, and $L_\text{m}$ is the size of the magnetic mirror, 
i.e., the variation scale of the mean magnetic field, the particle preserves its first adiabatic invariant, with a constant magnetic moment 
\begin{equation}
    \frac{m v_\perp^2 }{ 2B}= \text{constant}.
\end{equation}
Here $v_\perp$ is the particle velocity perpendicular to the magnetic field, $m$ is the particle mass, and $B$ is the magnetic field strength. 
When a particle with constant energy and magnetic moment
moves into a region with converging magnetic fields, 
its velocity along the magnetic field $v_\|$ decreases, and the angle between the particle velocity ${\bm v}$ and the magnetic field ${\bm B}$, i.e., 
the pitch angle, increases. 
The deceleration toward the mirror point where the mean magnetic field has the maximum strength and is 
perpendicular to the heliocentric radial direction
can potentially lead to the 
accumulation of the pickup ions with large pitch angles
\citep{Chal10}.
%and thus the enhanced ENA flux. 

% no trapping, not a bottle, deceleration happens for all initial pitch angles 

\begin{figure}[H]
\centering
   \includegraphics[width=9cm]{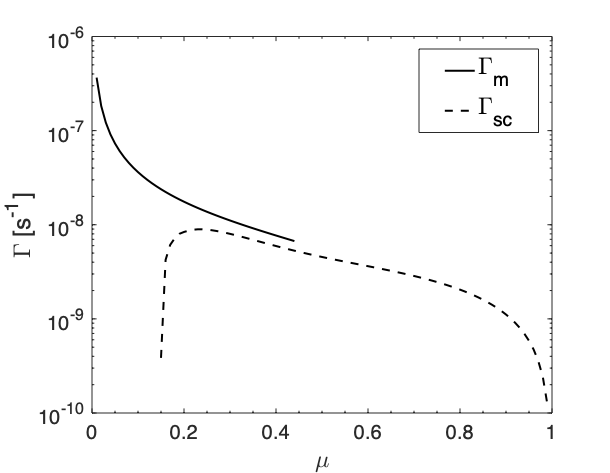}
\caption{The rate of mirroring $\Gamma_\text{m}$ by the interstellar mean magnetic field and the rate of scattering $\Gamma_\text{sc}$ by 
the turbulent magnetic fluctuations in the outer heliosheath for 
keV protons. 
The turbulence parameters are adopted (see text) to reach an approximate balance between $\Gamma_\text{m}$ and $\Gamma_\text{sc}$.
%$\delta B_f /B_0 =0.04$, $L =100$ au, $B_0 = 5~\mu$G
For the magnetic mirroring to be effective, the turbulence parameters should satisfy $\Gamma_\text{m}>\Gamma_\text{sc}$.}
\label{fig: meanmirror}
\end{figure}

The magnetic mirror force is 
\citep{CesK73}
\begin{equation}
    \frac{d (p \mu)}{dt} =  - \frac{v_\perp p_\perp}{2 B_0}   \frac{\delta B_\text{m}}{L_\text{m}},
\end{equation}
where $\mu$ is cosine of the pitch angle, $p$ is the particle momentum, $t$ is time, $B_0$ is the mean magnetic field strength, 
and $\delta B_\text{m}$ is the parallel magnetic fluctuation over $L_\text{m}$.
It follows that the rate of change in $\mu$ due to mirroring, i.e., the mirroring rate, is 
\begin{equation}
    \Gamma_\text{m} = \Big|\frac{1}{\mu} \frac{d\mu}{dt}\Big| =  \frac{v}{2B_0} \frac{\delta B_\text{m}}{L_\text{m}} \frac{1-\mu^2}{\mu},
\end{equation}
where $v$ is the particle speed. 
Under the consideration of keV protons ($v \approx 438$ km s$^{-1}$), $\delta B_\text{m} / B_0 \approx 1/4$ and $L_\text{m} \approx 100$ au
\citep{Chal10},
$\Gamma_\text{m}$ as a function of $\mu$ is presented in Fig. \ref{fig: meanmirror}.
It depends on the longitudinal gradient in the mean magnetic field. 
{The maximum $\mu$ for mirroring is determined by }
\begin{equation}
  \mu_\text{max} = \sqrt{\frac{\delta B_m}{B_0+\delta B_m}},
\end{equation}
which is derived from the first adiabatic invariant 
\begin{equation}
    \frac{v_\perp^2}{B_0} = \frac{v^2}{B_0+\delta B_m}.
\end{equation}

The turbulent magnetic fluctuations detected by {\it Voyager 1} and {\it 2}
in the outer heliosheath 
(e.g., \citealt{Burla18})
affects the transport of pickup protons 
via the field line wandering (see Section \ref{sec: wan}), 
turbulent mirroring (see Section \ref{sec: turmirr}), and pitch-angle scattering. 
As efficient scattering causes breaking of the first adiabatic invariant of particles, 
an effective magnetic mirroring imposes constraints on the amplitude of turbulent fluctuations.
Here we focus on the gyroresonant scattering 
{\citep{Jokipii1966,Kulsrud_Pearce,Schli02}}
of pickup protons by fast modes of MHD turbulence. 
Compared with Alfv\'{e}nic modes, fast modes usually have a smaller energy fraction 
\citep{CL02_PRL,Hu22cr}.
However, their isotropic energy scaling 
\citep{CL02_PRL}
leads to a more efficient gyroresonant scattering 
\citep{YL04,XY13,XL20}
compared to the anisotropic Alfv\'{e}nic modes
{\citep{Chan00,YL02,BYL2011}.}

The rate of scattering is defined as 
\citep{Jokipii1966,CesK73}
\begin{equation}\label{eq: gamsc}
   \Gamma_\text{sc}= \frac{1}{\mu^2} \Big\langle \frac{\delta \mu^2}{t} \Big\rangle = \frac{2 D_{\mu\mu}}{\mu^2},
\end{equation}
where $\delta \mu$ is the change of $\mu$ due to scattering, 
and $D_{\mu\mu}$ is the pitch-angle diffusion coefficient of fast modes
\citep{Volk:1975}
\begin{equation}\label{eq: duuvolk}
    D_{\mu\mu} = C_\mu \int d^3 k \frac{k_\|^2}{k^2} [J_1^\prime(x)]^2 I(k) R(k).
\end{equation}
In the above expression, 
{$J_1^\prime(x)$ is the
derivative of the Bessel function,}
and $I(k)$ is the magnetic energy spectrum of fast modes 
\citep{CL02_PRL},
\begin{equation}
    I(k) = C_f k^{-\frac{7}{2}}, ~~ C_f = \frac{1}{16\pi} \delta B_f^2 L^{-\frac{1}{2}},
\end{equation}
$k$ is the wavenumber, 
$L$ is the injection scale of turbulence, $\delta B_f$ is the rms strength of magnetic fluctuations of fast modes at $L$,
$R(k)$ is the resonance function for gyroresonance,
\begin{equation}
  R(k) = \pi \delta (\omega_k - v_\| k_\| + \Omega), ~~ \Omega = \frac{q B_0}{mc},
\end{equation}
$\Omega$ is the particle gyrofrequency, $\omega_k = k V_A $ is the wave frequency, 
$V_A =B_0/\sqrt{4\pi m_H n_i} $
is the Alfv\'{e}n speed in ions,
$m_H$ is the hydrogen atomic mass, 
$n_i$ is the ion number density,
$k_\|$ is the parallel component of $k$, 
$q$ is the electric charge, $m$ is the proton mass, $c$ is the light speed, 
\begin{equation}
   C_\mu = (1-\mu^2) \frac{\Omega^2}{B_0^2}, 
\end{equation}
\begin{equation}
   x = \frac{k_\perp v_\perp}{\Omega} = \frac{k_\perp}{r_g^{-1}}, ~~ r_g = \frac{v_\perp}{\Omega},
\end{equation}
and $k_\perp$ is the perpendicular component of $k$. 
The approximate expression of $\Gamma_\text{sc}$ in the limit of a small $x$ is 
\citep{XL20}, 
\begin{equation}\label{eq:fastsc}
    \Gamma_\text{sc} \approx \frac{\pi}{28} \frac{\delta B_f^2}{B_0^2} \Big(\frac{v}{L\Omega}\Big)^\frac{1}{2} \Omega (1-\mu^2)\mu^{-\frac{3}{2}},
\end{equation}
which depends on $\delta B_f$.

As discussed above, for the mirroring of the mean magnetic field to be effective, there should be $\Gamma_\text{sc} < \Gamma_\text{m}$.
In Fig. \ref{fig: meanmirror}, we present $\Gamma_\text{sc}$ for keV protons calculated using Eqs. \eqref{eq: gamsc} and \eqref{eq: duuvolk}, 
with $B_0 = 5~\mu$G,
$L = 100$ au, $\delta B_f / B_0 = 0.04$, and $n_i = 0.07$ cm$^{-3}$
\citep{Sla08,Swa20}. 
{We adopt $5\times10^7$ cm
as the cutoff scale of turbulent spectrum, which is comparable to the ion inertial length 
\citep{Frat21} and smaller than 
{$r_g \approx 9\times10^8$ cm at $\mu=0$}.
Due to the lack of turbulent magnetic fluctuations at small scales, 
the scattering is absent at small $\mu$. 
} 
Given the above parameters, the mirroring by the mean field and the scattering by turbulence 
are approximately in balance {over the $\mu$ values where they are both present.} 
The condition $\Gamma_\text{sc} < \Gamma_\text{m}$ can be used to constrain $\delta B_f$ and $L$ for the MHD turbulence 
in the outer heliosheath.

\section{Mirroring effect of turbulent magnetic fields}
\label{sec: turmirr}

The mirroring of the interstellar mean magnetic field alone may not be able to account for the ribbon. 
Moreover, 
\citet{Zirn18}
argued that the mean field mirror acts to push away pickup ions from the mirror point and reduces the ribbon flux.
In addition to the mirroring effect of the interstellar mean magnetic field, 
the compressible MHD turbulence in the outer heliosheath can also contribute to the magnetic mirroring. 
Both fast and slow modes create multi-scale magnetic mirrors within the inertial range of turbulence.
Isotropic fast modes are more efficient in mirroring than anisotropic slow modes
\citep{XL20}, 
as the latter have the magnetic fluctuations decrease 
more rapidly with decreasing scales in the direction parallel to the local magnetic field.

\begin{figure}[H]
\centering
   \includegraphics[width=9cm]{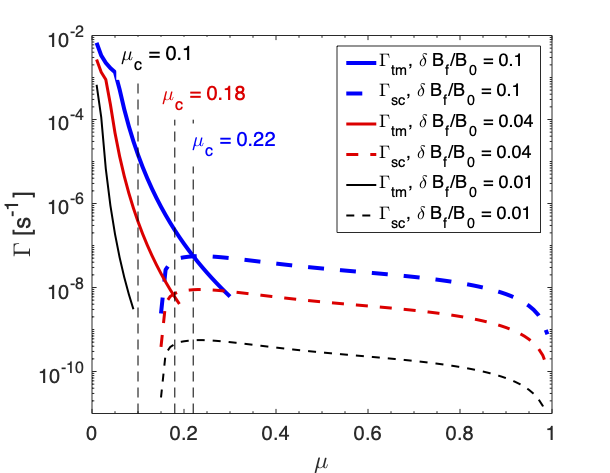}
\caption{The rate of mirroring $\Gamma_\text{tm}$ 
(Eqs. \eqref{eq: gamtmbi} and \eqref{eq: gamtmsm})
and the rate of scattering $\Gamma_\text{sc}$ (Eqs. \eqref{eq: gamsc} and \eqref{eq: duuvolk}) by 
the turbulent magnetic fluctuations in the outer heliosheath for 
keV protons. 
Same parameters as in Fig. \ref{fig: meanmirror} are used except for $\delta B_f /B_0$ values. 
%keV proton, $L = 100$ au, $B_0 = 5~\mu$G, 
The vertical dashed lines denote $\mu_c$.
%at the balance between $\Gamma_\text{tm}$ and $\Gamma_\text{sc}$.
}
\label{fig: tmdbf}
\end{figure}

Similar to the analysis in Section \ref{sec: mirrmean}, one easily finds that 
the mirroring rate of fast modes is 
\begin{equation}
   \Gamma_\text{tm} = \frac{v}{2B_0} b_{fk}k \frac{1-\mu^2}{\mu},
\end{equation}
where $b_{fk}$ is the magnetic fluctuation of fast modes at $k$.
Among the mirrors at different wavenumbers, the ones that are most effective in reflecting the particles at $\mu$ have 
\citep{CesK73}
\begin{equation}\label{eq: adiink}
  \mu^2 \approx \frac{b_{fk}}{B_0}.
\end{equation}
{Smaller mirrors would have insufficient magnetic fluctuation to reflect the particles, and larger mirrors would have a lower mirroring rate.}
Given the scaling relation of fast modes
\citep{CL02_PRL},
\begin{equation}\label{eq: fasvsc}
   b_{fk} = \delta B_f (kL)^{-\frac{1}{4}},
\end{equation}
there is \citep{XL20},
\begin{equation}\label{eq: gamtmbi}
    \Gamma_\text{tm} \approx \frac{v}{2L} \Big(\frac{\delta B_f}{B_0}\Big)^4 \frac{1-\mu^2}{\mu^7}. 
\end{equation}
We note that unlike the mirroring of the mean field, 
for the mirroring of turbulent magnetic fields, 
particles with a given $\mu$ predominantly interact with the mirrors at the corresponding $k$.
%With an increasing magnetic field gradient toward a larger $k$, 
$\Gamma_\text{tm}$ has a much stronger dependence on $\mu$ compared to $\Gamma_\text{m}$.
Despite the much smaller fluctuation of turbulent magnetic fields, 
$\Gamma_\text{tm}$ is significantly larger than $\Gamma_\text{m}$ due to the 
large magnetic field gradients at small scales
(see Figs. \ref{fig: meanmirror} and \ref{fig: tmdbf}).
It suggests that the mirroring effect of the turbulent magnetic fields dominates over that of the mean field.

Eq. \eqref{eq: gamtmbi} stands for the case with $r_g < 1/k < L$, that is  
(Eqs. \eqref{eq: adiink} and \eqref{eq: fasvsc})
\citep{XL20},
\begin{equation}
   \mu> \mu_{r_g} \approx \sqrt{\frac{b_{fk} (r_g)}{B_0}} =  \Big(\frac{\delta B_f}{B_0}\Big)^\frac{1}{2} \Big(\frac{r_g}{L}\Big)^\frac{1}{8},
\end{equation}
and 
\begin{equation}\label{eq:mumaxmc}
   \mu < \mu_\text{max} = \sqrt{\frac{\delta B_f}{B_0 + \delta B_f}},
\end{equation}
which is derived from the first adiabatic invariant, 
\begin{equation}
   \frac{v_\perp^2}{B_0} = \frac{v^2}{B_0 + \delta B_f}.
\end{equation}
For $1/k < r_g$, i.e., $\mu< \mu_{r_g}$, the mirrors at $1/k \approx r_g$ dominate the mirroring. Therefore, 
there is 
\citep{XL20}
\begin{equation}\label{eq: gamtmsm}
   \Gamma_\text{tm} = \frac{v}{2B_0} \frac{b_{fk}(r_g)}{r_g} \frac{1-\mu^2}{\mu} 
   = \frac{v}{2 r_g} \frac{\delta B_f}{B_0} \Big(\frac{r_g}{L}\Big)^\frac{1}{4} \frac{1-\mu^2}{\mu}. 
\end{equation}

In Fig. \ref{fig: tmdbf}, we present $\Gamma_\text{tm}$ and $\Gamma_\text{sc}$ for the mirroring and scattering of keV protons by 
fast modes. 
The same parameters as in Fig. \ref{fig: meanmirror} are used but for different $\delta B_f /B_0$ values. 
{If $\Gamma_\text{tm}$ and $\Gamma_\text{sc}$ can reach balance, }
the critical $\mu$, $\mu_c$, is defined at their balance.
With $\Gamma_\text{tm}> \Gamma_\text{sc}$ at $\mu< \mu_c$, 
the diffusion of protons (parallel to the local magnetic field) is dominated by mirroring (see Section \ref{sec: turdiff}). 
At larger $\mu$, scattering becomes more important than mirroring. 
As an approximation, by using Eqs. \eqref{eq:fastsc} and \eqref{eq: gamtmbi}, we find 
\citep{XL20}
\begin{equation}\label{eq: anamuc}
\begin{aligned}
  \mu_c &\approx \bigg[ \frac{14}{\pi} \frac{\delta B_f^2}{B_0^2} \Big(\frac{v}{L\Omega}\Big)^\frac{1}{2}\bigg]^\frac{2}{11} \\
             & \approx 0.11 \Big(\frac{\delta B_f/B_0}{0.04}\Big)^\frac{4}{11} \Big(\frac{L}{100 ~ \text{au}}\Big)^{-\frac{1}{11}} 
                \Big(\frac{B_0}{5 ~ \mu\text{G}}\Big)^{-\frac{1}{11}} \\
              &~~~~~ \Big(\frac{v}{438~\text{km s}^{-1}}\Big)^\frac{1}{11},
\end{aligned}
\end{equation}
which is close to the exact value of $\mu_c$ at the intersection between $\Gamma_\text{tm}$ and $\Gamma_\text{sc}$ (see Fig. \ref{fig: tmdbf}).
{The actual  $\mu_c$ is slightly larger than the above estimate due to the cutoff of turbulent spectrum that was not taken into account in 
\citet{XL20}.
If the balance between $\Gamma_\text{tm}$ and $\Gamma_\text{sc}$ cannot be reached (see the case with $\delta B_f /B_0 = 0.01$ in Fig. \ref{fig: tmdbf}), $\mu_c$ is determined by $\mu_\text{max}$ in Eq. \eqref{eq:mumaxmc}.}
By using Eqs. \eqref{eq: adiink} and \eqref{eq: fasvsc}, we find that the corresponding length scale is 
\begin{equation}\label{eq: kc}
\begin{aligned}
   k_c^{-1} &= L \Big(\frac{\delta B_f}{B_0}\Big)^{-4} \mu_c^8 \\
                 & \approx 43~\text{au} \Big(\frac{\delta B_f/B_0}{0.04}\Big)^{-4} \Big(\frac{L}{100~\text{au}}\Big)
                        \Big(\frac{\mu_c}{0.18}\Big)^8,
\end{aligned}
\end{equation}
as the maximum size of the mirrors in turbulent magnetic fields for mirroring keV protons, 
{which is much larger than $r_g$.}

% changing energy to 6 keV, mu_c doesn't change very much 

\section{Mirror diffusion and ribbon width}\label{sec: turdiff}

\begin{figure}[H]
\centering
   \includegraphics[width=8cm]{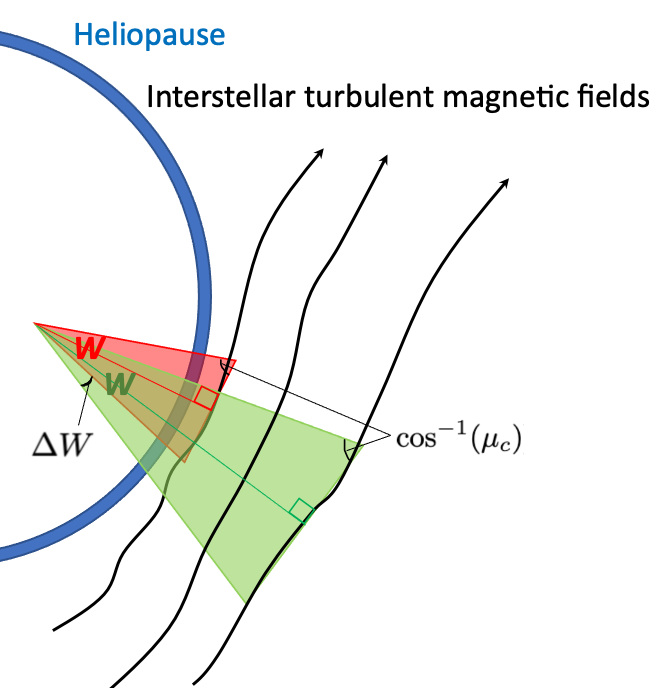}
\caption{Illustration for the ribbon width. {The shaded regions, after projection along the line of sight, correspond to the ENA flux distribution transverse to the ribbon structure in the sky.}
The width $W$ corresponds to 
the region in the outer heliosheath where pickup ions undergo the slow mirror diffusion in interstellar turbulent magnetic fields.  
The misalignment between red and green shaded regions shows the additional broadening of the ribbon $\Delta W$ due to the field line 
wandering caused by Alfv\'{e}nic modes over a radial distance 
$\sim \lambda$.
}
\label{fig: sket}
\end{figure}

{Based on the mirroring by fast modes and perpendicular superdiffusion and wandering of magnetic fields induced by Alfv\'{e}n modes, our model for interpreting the {\it IBEX} ribbon is illustrated in Fig. \ref{fig: sket}.}
Unlike the trapping of particles between the mirror points within static magnetic bottles, 
in {MHD turbulence with the perpendicular superdiffusion of magnetic fields
\citep{LV99,Eyin13},
particles that follow turbulent magnetic field lines also undergo the perpendicular superdiffusion
and thus cannot be spatially trapped} 
%As a result, they stochastically interact with different mirrors and exhibit the mirror diffusion along turbulent magnetic field lines
%\citep{LX21}.
%This mirror diffusion originates from the stochasticity of turbulent magnetic fields 
%\citep{LV99,Eyin13}.
%The accelerated growth of field line separation
%enabled by turbulent reconnection
%\citep{LV99}
%results in the 
%superdiffusion of CRs in the direction perpendicular to the mean field
\citep{XY13,LY14,Hu22cr}.
During the perpendicular superdiffusion, particles always encounter different mirrors, rather than bouncing back and forth between two mirror points. 
{The corresponding diffusion in the direction parallel to the magnetic field
is termed mirror diffusion in 
\citet{LX21}.}

It is known that the diffusion arising from pitch-angle scattering faces the so-called $90^\circ$ problem, 
that is, a particle cannot be scattered through $90^\circ$ 
(i.e., $\mu=0$) according to the quasi-linear theory
\citep{Jokipii1966}
(see Fig. \ref{fig: tmdbf}),  
leading to an infinitely large parallel mean free path
(however, see e.g., 
\citet{Xuc16,XLb18}
for gyroresonant scattering with turbulence-broadened resonance).
The mirror diffusion does not depend on the pitch angle scattering. 
For a particle with a given $\mu$, the step size of its random walk along the magnetic field is the size of the mirrors that are most effective in 
reflecting the particle (see Section \ref{sec: turmirr}).
Therefore, its spatial diffusion coefficient along the turbulent field line is 
(Eqs. \eqref{eq: adiink} and \eqref{eq: fasvsc})
\citep{LX21}
\begin{equation}\label{eq: dpamul}
   D_\|(\mu) \approx v\mu k^{-1} = v L \Big(\frac{\delta B_f}{B_0}\Big)^{-4} \mu^9, ~~  \mu_{r_g} < \mu < \mu_c,  
\end{equation}
which decreases drastically with decreasing $\mu$.
For $\mu < \mu_{r_g}$, there is 
\begin{equation}\label{eq: dpamus}
   D_\|(\mu) \approx v \mu r_g, ~~ \mu < \mu_{r_g}. 
\end{equation}

Near the region with ${\bm B} \cdot {\bm r} = 0$, where ${\bm r}$ is the radial line of sight direction, 
%zero radial component of the interstellar mean magnetic field,
the pickup ions from neutral solar wind
\citep{Zim19}
are subject to the mirroring effect of the mean field, 
provided that the turbulent magnetic fluctuations are sufficiently low for the mirroring to dominate over the turbulent scattering
(see Section \ref{sec: mirrmean}).
The pitch-angle distribution of the pickup ions is expected to become more anisotropic toward the mirror point, 
with an excess at large pitch angles. 
The pickup ions accumulated near the mirror point of the mean field with $\mu < \mu_c$ 
and the pickup ions with initial $\mu$'s less than $\mu_c$ 
can interact with the mirrors in turbulent magnetic fields (see Section \ref{sec: turmirr}) and undergo the 
mirror diffusion.
The largest diffusion distance over their $\sim1$ year lifetime 
\citep{Gia15}
can be estimated as 
\begin{equation}\label{eq:diffdis}
\begin{aligned}
   S_D(\mu)&=\sqrt{2 D_\| (\mu) t}  \\
   &\approx  38~\text{au}
   \Big(\frac{\delta B_f/B_0}{0.04}\Big)^{-2} \Big(\frac{L}{100~\text{au}}\Big)^\frac{1}{2} \\
   &~~~~~\Big(\frac{v}{438~\text{km s}^{-1}}\Big)^\frac{1}{2} 
    \Big(\frac{\mu}{0.18}\Big)^\frac{9}{2} 
   \Big(\frac{t}{1~\text{year}}\Big)^\frac{1}{2}.
\end{aligned}
\end{equation}
%which is much smaller than the particle streaming distance along the magnetic field, 
%\begin{equation}
%   v\mu t \approx  9.2~\text{au}
%   \Big(\frac{v}{438~\text{km s}^{-1}}\Big)  \Big(\frac{\mu}{0.1}\Big)  \Big(\frac{t}{1~\text{year}}\Big).
%\end{equation}
With the strong dependence of $S_D(\mu)$ on $\mu$,
the pickup ions with smaller $\mu$'s have much more suppressed diffusion and are accumulated 
near the point with ${\bm B} \cdot {\bm r} = 0$, causing the more enhanced ENA flux 
toward the center of the ribbon. 
With $\Gamma_m \ll \Gamma_\text{tm}$
(see Section \ref{sec: turmirr}),
the effect of the mean field mirror on reflecting particles away from the mirror point 
\citep{Zirn18}
is insignificant.

The width of the ribbon is determined by $\mu_c$, 
\begin{equation}\label{eq: fastrw}
    W \approx  2[90^\circ - \cos^{-1} (\mu_c)].
\end{equation}
{With $\mu_c$ approximately proportional to $(\delta B_f /B_0)^{4/11}$ (Eq. \eqref{eq: anamuc}),  
$W$ increases with $\delta B_f$.}
For $\mu_c = 0.18$ 
in the case with $\delta B_f /B_0 = 0.04$
(see Fig. \ref{fig: tmdbf}), the corresponding $W \approx 21^\circ$
is consistent with the observed ribbon width 
of $\sim 20^\circ$ at $\sim 1$ keV
\citep{Schw11}.
The observed ribbon width provides a constraint on the maximum $\delta B_f$.

%Given the ribbon width $\sim 20^\circ$ and the ribbon source distance $\approx 150$ au
%\citep{Zir20}, 
%the size of the ribbon source 
%in the transverse direction (i.e., parallel to the interstellar magnetic field) is $\approx 53$ au.
%It is much larger than $S_D(\mu)$, showing that the pickup ions are well spatially confined 
%due to the mirror diffusion
%during their lifetime. 

As shown in Fig. \ref{fig: sket},
the shaded region indicates the ribbon width $W$.
The pickup ions with initial $\mu$'s smaller than $\mu_c$ are subject to the mirroring of the interstellar turbulent magnetic fields and undergo the {$\mu$-dependent mirror diffusion}, resulting in the 
enhanced pickup-ion intensity.
The pickup ions with larger initial $\mu$'s are scattered by interstellar turbulent magnetic fields.
The diffusion associated with scattering is much faster than the mirror diffusion 
\citep{LX21}, 
and the particles lose their 
memory of the initial pitch angles. 
Therefore, the pickup ions with initial $\mu$'s larger than $\mu_c$, after being neutralized, are unlikely to 
return to the inner heliosphere to be observed by {\it IBEX}.
We note that as $\delta B_f$ of fast modes may change with the distance from the heliopause, $W$ corresponding to different distances beyond the heliopause can be different. 
{Future observations by
Interstellar Mapping and Acceleration Probe (IMAP, \citealt{imap16})
that will resolve the substructure of the ribbon
might provide more information on the variation of $W$ along the line of sight to test our theory.}

\section{The effect of field line wandering on the ribbon}
\label{sec: wan}

The confinement of pickup ions near the region with ${\bm B} \cdot {\bm r} = 0$ is attributed to their mirroring interaction with the
compressible modes of MHD turbulence on length scales less than $k_c^{-1}$ (Eq. \eqref{eq: kc}).
The incompressible Alfv\'{e}nic modes do not contribute to mirroring and have a negligible contribution to scattering due to their anisotropy, 
but they dominate the wandering of turbulent magnetic field lines
\citep{LV99}. 
Next we will discuss the effect of field line wandering on 
broadening of the ribbon and distortion of the ribbon structure.

The ribbon source region stretches out from the heliopause to a radial distance $\sim 100$ au that is determined by the 
mean free path $\lambda$ of neutral solar wind atoms beyond the heliopause 
\citep{Hee16}.
We extrapolate the observed magnetic fluctuation $\delta B_\text{obs} \approx 0.3~\mu$G at length scale $l_\text{obs} \approx 20$ au
by Voyager 1 in the VLISM 
\citep{Lee20}
by following the Kolmogorov scaling 
\citep{Burla18,Lee20} 
and find the magnetic fluctuation at length scale $\lambda$,
\begin{equation}
\begin{aligned}
   \delta B_\lambda &= \delta B_\text{obs} \Big(\frac{l_\text{obs}}{\lambda}\Big)^{-\frac{1}{3}} \\
    &\approx  0.5~\mu\text{G}\Big(\frac{\delta B_\text{obs}}{0.3~\mu\text{G}}\Big) \Big(\frac{l_\text{obs}}{20~\text{au}}\Big)^{-\frac{1}{3}}
    \Big(\frac{\lambda}{100~\text{au}}\Big)^\frac{1}{3}.
\end{aligned}
\end{equation}
We assume that the measured magnetic fluctuation is mainly induced by Alfv\'{e}nic modes based on the observed Kolmogorov magnetic energy spectrum. 
Different from the mirroring effect of compressible fast and slow modes,
incompressible Alfv\'{e}nic modes cause wandering of magnetic field lines. 
The degree of wandering over a perpendicular length scale $\lambda$ is determined by the corresponding magnetic fluctuation, with 
\begin{equation}
    \tan \theta_\lambda = \frac{\delta B_\lambda}{B_0}.
\end{equation}
%where $\theta$ is the angle between ${\bm B_0}$ and total magnetic field ${\bm B}$. 
The wandering of field lines within the ribbon source causes shift of the points with ${\bm B} \cdot {\bm r} = 0$
and thus shift of the ribbon center
(see Fig.~\ref{fig: sket}).
Given the above estimate of $\delta B_\lambda$ and $B_0 = 5~\mu$G, we find that the additional broadening of the ribbon 
due to the field line wandering along the line of sight
\begin{equation}\label{eq: alfvbro}
   \Delta W = \tan^{-1}\Big(\frac{\delta B_\lambda}{B_0}\Big) \approx 5.7^\circ,
\end{equation}
{which provides the upper limit for the broadening caused by Alfv\'{e}nic modes.}
%superposition of shifted
Combining the broadening effects of both Alfv\'{e}nic and fast modes, we have the ribbon width 
(Eqs. \eqref{eq: fastrw} and \eqref{eq: alfvbro})
\begin{equation}\label{eq: combwid}
   W + \Delta W \approx  2 [90^\circ - \cos^{-1}(\mu_c)] + \tan^{-1} \Big(\frac{\delta B_\lambda}{B_0}\Big).
\end{equation}
%which is approximately $23^\circ$ for $\mu_c = 0.15$ and $\lambda = 100$ au.
{Both broadening effects are illustrated in Fig.~\ref{fig: sket}.

In addition to the above broadening effect,}
the field line wandering induced by Alfv\'{e}nic modes can also cause distortion of the ribbon structure across the sky.
The turbulent magnetic fluctuation at $L$ in the VLISM is 
\begin{equation}\label{eq: debinjalf}
   \delta B_L = \delta B_\text{obs} \Big(\frac{l_\text{obs}}{L}\Big)^{-\frac{1}{3}}.
\end{equation}
Under the assumption that $\delta B_L$ is mainly associated with Alfv\'{e}nic modes, the field line wandering over $L$ 
in the direction perpendicular to the radial line of sight and perpendicular to the interstellar mean magnetic field
can give rise to the misalignment of ribbon centers across the sky by $\theta_L$, with 
\begin{equation}
    \tan \theta_L = \frac{\delta B_L}{B_0}.
\end{equation}
For the ribbon structure to remain spatially coherent across the sky, $\theta_L$ should be smaller than half of the average ribbon width, i.e., 
$\theta_L < 10^\circ$.
This constraint leads to 
\begin{equation}
   \delta B_L < 0.88~\mu\text{G} \Big(\frac{\tan\theta_L}{\tan(10^\circ)}\Big)\Big(\frac{B_0}{5~\mu\text{G}}\Big),
\end{equation}
and (Eq. \eqref{eq: debinjalf})
\begin{equation}\label{eq: newlinj}
\begin{aligned}
L <   508~\text{au}   
      \Big(\frac{\tan\theta_L}{\tan(10^\circ)}\Big)^3 \Big(\frac{\delta B_\text{obs}}{0.3~\mu\text{G}}\Big)^{-3} 
      \Big(\frac{B_0}{5~\mu\text{G}}\Big)^3
              \Big(\frac{l_\text{obs}}{20~\text{au}}\Big).
\end{aligned}
\end{equation}
The above estimate is consistent with the simulation in 
\citet{Zir20}, 
where they found that given $L = 500$ au, 
the ribbon at large scales changes drastically, with the peak of the ribbon meandering around the sky.

In \citet{XLi22}, an upper limit of $L$ was found to be $\sim 194$ au
under the ion-neutral decoupling condition in the partially ionized VLISM.
It is more stringent than that imposed by the spatial coherence of the ribbon (Eq. \eqref{eq: newlinj}). 
%If we adopt $L \approx 100$ au, the corresponding $\delta B_L$ is $\approx 0.5~\mu$G (Eq. \eqref{eq: debinjalf}).
%The upper limit of $\delta B_f$ constrained by the ribbon width is $\approx 0.04 B_0\approx 0.2~\mu$G (Eq. \eqref{eq: fastrw}), 
%which is smaller than $\delta B_L$.
%The relation $\delta B_f < \delta B_L$ is supported by the observational finding on dominant transverse magnetic fluctuations detected 
%more recently by {\it Voyager} 1 
%\citep{Burla18}.

\section{Discussion}

{Earlier studies on 
magnetic mirroring 
\citep{CesK73} and 
its application to trapping of 
pickup ions 
\citep{Gia15,Zir20}
adopt a model of MHD turbulence as a superposition of MHD waves. 
%As MHD waves only have oscillatory motions and are fixed in space, magnetic fields in such models do not undergo the turbulent reconnection
%\citep{LV99}
%and superdiffusion in the direction perpendicular to the mean magnetic field 
%\citep{LVC04,Beresd13}. 
%Accordingly, particles are able to repeatedly trace back the same field line and spatially trapped between two fixed magnetic mirror points. 
In the MHD turbulence model with strong nonlinear interactions between oppositely directed wave packets, 
the turbulent energy cascade mainly happens in the direction perpendicular to the local magnetic field
\citep{GS95,LV99,CL02_PRL}.
The anisotropic nature of MHD turbulence was not taken into account
in 
\citet{Gia15,Zir20}, 
so it is unclear what types of magnetic fluctuations are most efficient in interacting with the $\sim$ keV ions in their model.
In nonlinear MHD turbulence, due to the stochasticity and perpendicular superdiffusion of turbulent magnetic fields
\citep{Eyink2011,BL19}, particles simultaneously undergo the perpendicular superdiffusion 
\citep{XY13,LY14,Hu22cr,LXH23}
by following the turbulent field lines and mirror reflection by magnetic compressions
\citep{ZX23}. 
The former process enables encounters of a particle with multiple magnetic mirrors and its diffusive motion along the magnetic field. Therefore, in both parallel and perpendicular directions, we expect that particles are not trapped.
More detailed comparison with earlier studies will be carried out by using test particle simulations in MHD turbulence
in our future work. }

{In addition, the relation between the characteristic MHD turbulence parameters, e.g., the energy fraction and scaling of compressible modes, the injection scale of turbulence, and the ribbon width was not clearly established in earlier studies. 
In this work,} 
we quantified the range of pitch angles for effective turbulent mirroring with $\mu < \mu_c$, where $\mu_c$ is determined by the balance between mirroring and scattering (Eq. \eqref{eq: anamuc})
or by $\mu_\text{max}$ (Eq. \eqref{eq:mumaxmc}),  
the corresponding range of scales 
$r_g < k^{-1} < k_c^{-1}$ (Eq. \eqref{eq: kc}), 
and the diffusion coefficient and distance 
of mirror diffusion 
(Eqs. \eqref{eq: dpamul}, \eqref{eq: dpamus}, and \eqref{eq:diffdis}).
The ribbon width depends on $\mu_c$ and thus can provide a constraint on the magnetic fluctuation of compressible modes, 
{whose energy fraction may change with the radial distance away from the heliopause.
Analysis of {\it Voyager 1} data reveals that the magnetic fluctuations in the VLISM are of mixed compressible and transverse nature
\citep{Burla18,Frat21}.
Studies by, e.g., 
\citet{Zank17,Matsu20}
suggest that the turbulent fluctuations in the solar wind are transmitted into the VLISM as fast modes. 
A significant level of 
compressible magnetic fluctuations is still seen in the VLISM in late 2018
\citep{Frat21}. Based on these studies, we assume that magnetic compressions are present in an extended region spanning over $\sim 100$ au beyond the heliopause and account for the turbulent mirroring.}
The mirror diffusion is sufficiently slow to account for the 
confinement of pickup ions near the point with ${\bm B} \cdot {\bm r} = 0$. 
{In addition, it preserves the initial pitch angles of particles for the ENAs to travel back to 1 au and generate the ribbon. }

With $\lambda$ of neutral atoms increasing with increasing energies
\citep{Lind05},
the ribbon source region extends farther into the VLISM at higher energies
\citep{Zirn16}.
With more significant field line wandering within the ribbon source, a broader ribbon is expected at higher energies. 
This is consistent with the observed ribbon width that increases with increasing ENA energies
\citep{Schwa11}.
The multi-scale turbulent motions of magnetic field lines, including both turbulent compression of magnetic fields and field line wandering, 
can also give rise to fine structure seen in the ribbon 
\citep{Mcc09}
and temporal variations.

{We note that the turbulent spectrum in the VLISM on scales less than $\sim 6 \times10^{11}$ cm 
becomes shallower than the Kolmogorov scaling
\citep{Lee20}. 
Despite the unclear physical origin and large data uncertainty
\citep{Frat21}, 
the enhanced small-scale magnetic fluctuations can affect both mirroring and scattering. }

We did not consider 
pickup-ion generated waves
(e.g., \citealt{SchMc13}).
Although the enhanced scattering by self-generated waves can suppress the spatial diffusion of pickup ions, 
strong scattering also causes significant changes in pitch angles. 
Pitch-angle isotropization by strong scattering 
near ${\bm B} \cdot {\bm r} = 0$ would reduce the ribbon flux. 

We did not consider the draping of the large-scale interstellar magnetic field around the heliosphere, which can also act to broaden the ribbon 
\citep{Zir20}.

%didn't consider phase speed close to particle parallel speed. 

\section{Summary}

To investigate the origin of the {\it IBEX} ribbon,
we studied 
%the mirroring effect on confining 
the interaction of 
pickup ions 
with MHD turbulence in the VLISM. 
%with large pitch angles 
%near ${\bm B} \cdot {\bm r} = 0$ in the outer heliosheath.
The magnetic compressions in the MHD turbulence serve as magnetic mirrors. The mirroring effect of turbulent magnetic fields dominates over that of compressed mean magnetic field due to the large turbulent magnetic field gradient at small scales. 
{Due to the perpendicular superdiffusion of turbulent magnetic fields induced by Alfv\'{e}nic modes,
particles interacting with turbulent magnetic mirrors undergo mirror diffusion parallel to the magnetic field.} 
The mirror diffusion of pickup ions is sufficiently slow,
causing their accumulation near ${\bm B} \cdot {\bm r} = 0$.
{With the initial pitch angles preserved during the mirror diffusion, 
after the pickup ions are neutralized, the resulting 
secondary ENAs can  
return to the heliosphere to be observed by {\it IBEX}.}

The smallest pitch angle for effective mirroring is determined by the balance between turbulent mirroring and pitch-angle scattering 
{and the amplitude of magnetic compression}. 
The ribbon width depends on the range of pitch angles for effective turbulent mirroring (see Eq. \eqref{eq: fastrw}). 
It provides a constraint on the magnetic fluctuation of fast modes that dominate both mirroring and scattering.

The field line wandering caused by Alfv\'{e}nic modes
can further broaden the ribbon by shifting the 
locations with ${\bm B} \cdot {\bm r} = 0$ within the ribbon source region (see Eq. \eqref{eq: combwid}).
For the ribbon structure to be coherent across the sky, 
under the assumption that the MHD turbulence 
measured by {\it Voyager} 1 in the VLISM is dominated by Alfv\'{e}nic modes, we find that 
the injection scale of turbulence is limited up to $\sim 500$ au (Eq. \eqref{eq: newlinj}). 
This is larger than the upper limit ($\sim 194$ au) imposed by ion-neutral decoupling condition 
in the partially ionized VLISM 
\citep{XLi22}.

\acknowledgments
S.X. acknowledges  
the support for 
this work provided by NASA through the NASA Hubble Fellowship grant \# HST-HF2-51473.001-A awarded by the Space Telescope Science Institute, which is operated by the Association of Universities for Research in Astronomy, Incorporated, under NASA contract NAS5-26555. 
H.L. acknowledges 
the support by LANL LDRD program and DOE OFES program. 
H.L. gratefully acknowledges discussions with Fan Guo.
\software{MATLAB \citep{MATLAB:2021}}

\bibliographystyle{aasjournal}
\bibliography{xu}

\begin{thebibliography}{}
\expandafter\ifx\csname natexlab\endcsname\relax\def\natexlab#1{#1}\fi
\providecommand{\url}[1]{\href{#1}{#1}}
\providecommand{\dodoi}[1]{doi:~\href{http://doi.org/#1}{\nolinkurl{#1}}}
\providecommand{\doeprint}[1]{\href{http://ascl.net/#1}{\nolinkurl{http://ascl.net/#1}}}
\providecommand{\doarXiv}[1]{\href{https://arxiv.org/abs/#1}{\nolinkurl{https://arxiv.org/abs/#1}}}

\bibitem[{{Beresnyak}(2013)}]{Beresd13}
{Beresnyak}, A. 2013, \apjl, 767, L39, \dodoi{10.1088/2041-8205/767/2/L39}

\bibitem[{{Beresnyak} \& {Lazarian}(2019)}]{BL19}
{Beresnyak}, A., \& {Lazarian}, A. 2019, {Turbulence in Magnetohydrodynamics}

\bibitem[{{Beresnyak} {et~al.}(2011){Beresnyak}, {Yan}, \&
  {Lazarian}}]{BYL2011}
{Beresnyak}, A., {Yan}, H., \& {Lazarian}, A. 2011, \apj, 728, 60,
  \dodoi{10.1088/0004-637X/728/1/60}

\bibitem[{{Burlaga} {et~al.}(2018){Burlaga}, {Florinski}, \& {Ness}}]{Burla18}
{Burlaga}, L.~F., {Florinski}, V., \& {Ness}, N.~F. 2018, \apj, 854, 20,
  \dodoi{10.3847/1538-4357/aaa45a}

\bibitem[{{Burlaga} {et~al.}(2022){Burlaga}, {Ness}, {Berdichevsky}, {Jian},
  {Kurth}, {Park}, {Rankin}, \& {Szabo}}]{Bur22}
{Burlaga}, L.~F., {Ness}, N.~F., {Berdichevsky}, D.~B., {et~al.} 2022, \apj,
  932, 59, \dodoi{10.3847/1538-4357/ac658e}

\bibitem[{{Cesarsky} \& {Kulsrud}(1973)}]{CesK73}
{Cesarsky}, C.~J., \& {Kulsrud}, R.~M. 1973, ApJ, 185, 153

\bibitem[{{Chalov} {et~al.}(2010){Chalov}, {Alexashov}, {McComas}, {Izmodenov},
  {Malama}, \& {Schwadron}}]{Chal10}
{Chalov}, S.~V., {Alexashov}, D.~B., {McComas}, D., {et~al.} 2010, \apjl, 716,
  L99, \dodoi{10.1088/2041-8205/716/2/L99}

\bibitem[{{Chandran}(2000)}]{Chan00}
{Chandran}, B.~D.~G. 2000, Physical Review Letters, 85, 4656,
  \dodoi{10.1103/PhysRevLett.85.4656}

\bibitem[{{Cho} \& {Lazarian}(2002)}]{CL02_PRL}
{Cho}, J., \& {Lazarian}, A. 2002, Physical Review Letters, 88, 245001,
  \dodoi{10.1103/PhysRevLett.88.245001}

\bibitem[{{Cho} \& {Lazarian}(2003)}]{CL03}
---. 2003, \mnras, 345, 325, \dodoi{10.1046/j.1365-8711.2003.06941.x}

\bibitem[{{Eyink} {et~al.}(2013){Eyink}, {Vishniac}, {Lalescu}, {Aluie},
  {Kanov}, {B{\"u}rger}, {Burns}, {Meneveau}, \& {Szalay}}]{Eyin13}
{Eyink}, G., {Vishniac}, E., {Lalescu}, C., {et~al.} 2013, \nat, 497, 466,
  \dodoi{10.1038/nature12128}

\bibitem[{{Eyink} {et~al.}(2011){Eyink}, {Lazarian}, \& {Vishniac}}]{Eyink2011}
{Eyink}, G.~L., {Lazarian}, A., \& {Vishniac}, E.~T. 2011, \apj, 743, 51,
  \dodoi{10.1088/0004-637X/743/1/51}

\bibitem[{{Fraternale} \& {Pogorelov}(2021)}]{Frat21}
{Fraternale}, F., \& {Pogorelov}, N.~V. 2021, \apj, 906, 75,
  \dodoi{10.3847/1538-4357/abc88a}

\bibitem[{{Giacalone} \& {Jokipii}(1999)}]{Giacalone_Jok1999}
{Giacalone}, J., \& {Jokipii}, J.~R. 1999, \apj, 520, 204,
  \dodoi{10.1086/307452}

\bibitem[{{Giacalone} \& {Jokipii}(2015)}]{Gia15}
---. 2015, \apjl, 812, L9, \dodoi{10.1088/2041-8205/812/1/L9}

\bibitem[{{Goldreich} \& {Sridhar}(1995)}]{GS95}
{Goldreich}, P., \& {Sridhar}, S. 1995, \apj, 438, 763, \dodoi{10.1086/175121}

\bibitem[{{Heerikhuisen} {et~al.}(2016){Heerikhuisen}, {Gamayunov},
  {Zirnstein}, \& {Pogorelov}}]{Hee16}
{Heerikhuisen}, J., {Gamayunov}, K.~V., {Zirnstein}, E.~J., \& {Pogorelov},
  N.~V. 2016, \apj, 831, 137, \dodoi{10.3847/0004-637X/831/2/137}

\bibitem[{{Hu} {et~al.}(2022){Hu}, {Lazarian}, \& {Xu}}]{Hu22cr}
{Hu}, Y., {Lazarian}, A., \& {Xu}, S. 2022, \mnras, 512, 2111,
  \dodoi{10.1093/mnras/stac319}

\bibitem[{{Jokipii}(1966)}]{Jokipii1966}
{Jokipii}, J.~R. 1966, \apj, 146, 480, \dodoi{10.1086/148912}

\bibitem[{{Kulsrud} \& {Pearce}(1969)}]{Kulsrud_Pearce}
{Kulsrud}, R., \& {Pearce}, W.~P. 1969, \apj, 156, 445, \dodoi{10.1086/149981}

\bibitem[{{Lazarian} \& {Vishniac}(1999)}]{LV99}
{Lazarian}, A., \& {Vishniac}, E.~T. 1999, \apj, 517, 700,
  \dodoi{10.1086/307233}

\bibitem[{{Lazarian} \& {Xu}(2021)}]{LX21}
{Lazarian}, A., \& {Xu}, S. 2021, \apj, 923, 53,
  \dodoi{10.3847/1538-4357/ac2de9}

\bibitem[{{Lazarian} {et~al.}(2023){Lazarian}, {Xu}, \& {Hu}}]{LXH23}
{Lazarian}, A., {Xu}, S., \& {Hu}, Y. 2023, Frontiers in Astronomy and Space
  Sciences, 10, 1154760, \dodoi{10.3389/fspas.2023.1154760}

\bibitem[{{Lazarian} \& {Yan}(2014)}]{LY14}
{Lazarian}, A., \& {Yan}, H. 2014, \apj, 784, 38,
  \dodoi{10.1088/0004-637X/784/1/38}

\bibitem[{{Lee} \& {Lee}(2020)}]{Lee20}
{Lee}, K.~H., \& {Lee}, L.~C. 2020, \apj, 904, 66,
  \dodoi{10.3847/1538-4357/abba20}

\bibitem[{{Lindsay} \& {Stebbings}(2005)}]{Lind05}
{Lindsay}, B.~G., \& {Stebbings}, R.~F. 2005, Journal of Geophysical Research
  (Space Physics), 110, A12213, \dodoi{10.1029/2005JA011298}

\bibitem[{MATLAB(2021)}]{MATLAB:2021}
MATLAB. 2021, MATLAB and Statistics Toolbox Release 2021b (Natick,
  Massachusetts: The MathWorks Inc.)

\bibitem[{{Matsukiyo} {et~al.}(2020){Matsukiyo}, {Noumi}, {Zank}, {Washimi}, \&
  {Hada}}]{Matsu20}
{Matsukiyo}, S., {Noumi}, T., {Zank}, G.~P., {Washimi}, H., \& {Hada}, T. 2020,
  \apj, 888, 11, \dodoi{10.3847/1538-4357/ab54c9}

\bibitem[{{McComas} {et~al.}(2009){McComas}, {Allegrini}, {Bochsler},
  {Bzowski}, {Christian}, {Crew}, {DeMajistre}, {Fahr}, {Fichtner}, {Frisch},
  {Funsten}, {Fuselier}, {Gloeckler}, {Gruntman}, {Heerikhuisen}, {Izmodenov},
  {Janzen}, {Knappenberger}, {Krimigis}, {Kucharek}, {Lee}, {Livadiotis},
  {Livi}, {MacDowall}, {Mitchell}, {M{\"o}bius}, {Moore}, {Pogorelov},
  {Reisenfeld}, {Roelof}, {Saul}, {Schwadron}, {Valek}, {Vanderspek}, {Wurz},
  \& {Zank}}]{Mcc09}
{McComas}, D.~J., {Allegrini}, F., {Bochsler}, P., {et~al.} 2009, Science, 326,
  959, \dodoi{10.1126/science.1180906}

\bibitem[{{Ocker} {et~al.}(2021){Ocker}, {Cordes}, {Chatterjee}, {Gurnett},
  {Kurth}, \& {Spangler}}]{Ock21}
{Ocker}, S.~K., {Cordes}, J.~M., {Chatterjee}, S., {et~al.} 2021, Nature
  Astronomy, 5, 761, \dodoi{10.1038/s41550-021-01363-7}

\bibitem[{{Pogorelov} {et~al.}(2011){Pogorelov}, {Heerikhuisen}, {Zank},
  {Borovikov}, {Frisch}, \& {McComas}}]{PogH11}
{Pogorelov}, N.~V., {Heerikhuisen}, J., {Zank}, G.~P., {et~al.} 2011, \apj,
  742, 104, \dodoi{10.1088/0004-637X/742/2/104}

\bibitem[{{Schlickeiser}(2002)}]{Schli02}
{Schlickeiser}, R. 2002, {Cosmic Ray Astrophysics}

\bibitem[{{Schwadron} \& {McComas}(2013)}]{SchMc13}
{Schwadron}, N.~A., \& {McComas}, D.~J. 2013, \apj, 764, 92,
  \dodoi{10.1088/0004-637X/764/1/92}

\bibitem[{{Schwadron} {et~al.}(2016){Schwadron}, {Opher}, {Kasper}, {Mewaldt},
  {Moebius}, {Spence}, \& {Zurbuchen}}]{imap16}
{Schwadron}, N.~A., {Opher}, M., {Kasper}, J., {et~al.} 2016, in Journal of
  Physics Conference Series, Vol. 767, Journal of Physics Conference Series,
  012025, \dodoi{10.1088/1742-6596/767/1/012025}

\bibitem[{{Schwadron} {et~al.}(2009){Schwadron}, {Bzowski}, {Crew}, {Gruntman},
  {Fahr}, {Fichtner}, {Frisch}, {Funsten}, {Fuselier}, {Heerikhuisen},
  {Izmodenov}, {Kucharek}, {Lee}, {Livadiotis}, {McComas}, {Moebius}, {Moore},
  {Mukherjee}, {Pogorelov}, {Prested}, {Reisenfeld}, {Roelof}, \&
  {Zank}}]{Schwad09}
{Schwadron}, N.~A., {Bzowski}, M., {Crew}, G.~B., {et~al.} 2009, Science, 326,
  966, \dodoi{10.1126/science.1180986}

\bibitem[{{Schwadron} {et~al.}(2011{\natexlab{a}}){Schwadron}, {Allegrini},
  {Bzowski}, {Christian}, {Crew}, {Dayeh}, {DeMajistre}, {Frisch}, {Funsten},
  {Fuselier}, {Goodrich}, {Gruntman}, {Janzen}, {Kucharek}, {Livadiotis},
  {McComas}, {Moebius}, {Prested}, {Reisenfeld}, {Reno}, {Roelof}, {Siegel}, \&
  {Vanderspek}}]{Schw11}
{Schwadron}, N.~A., {Allegrini}, F., {Bzowski}, M., {et~al.}
  2011{\natexlab{a}}, \apj, 731, 56, \dodoi{10.1088/0004-637X/731/1/56}

\bibitem[{{Schwadron} {et~al.}(2011{\natexlab{b}}){Schwadron}, {Allegrini},
  {Bzowski}, {Christian}, {Crew}, {Dayeh}, {DeMajistre}, {Frisch}, {Funsten},
  {Fuselier}, {Goodrich}, {Gruntman}, {Janzen}, {Kucharek}, {Livadiotis},
  {McComas}, {Moebius}, {Prested}, {Reisenfeld}, {Reno}, {Roelof}, {Siegel}, \&
  {Vanderspek}}]{Schwa11}
---. 2011{\natexlab{b}}, \apj, 731, 56, \dodoi{10.1088/0004-637X/731/1/56}

\bibitem[{{Slavin} \& {Frisch}(2008)}]{Sla08}
{Slavin}, J.~D., \& {Frisch}, P.~C. 2008, \aap, 491, 53,
  \dodoi{10.1051/0004-6361:20078101}

\bibitem[{{Swaczyna} {et~al.}(2020){Swaczyna}, {McComas}, {Zirnstein},
  {Sok{\'o}{\l}}, {Elliott}, {Bzowski}, {Kubiak}, {Richardson},
  {Kowalska-Leszczynska}, {Stern}, {Weaver}, {Olkin}, {Singer}, \&
  {Spencer}}]{Swa20}
{Swaczyna}, P., {McComas}, D.~J., {Zirnstein}, E.~J., {et~al.} 2020, \apj, 903,
  48, \dodoi{10.3847/1538-4357/abb80a}

\bibitem[{{Voelk}(1975)}]{Volk:1975}
{Voelk}, H.~J. 1975, Reviews of Geophysics and Space Physics, 13, 547,
  \dodoi{10.1029/RG013i004p00547}

\bibitem[{{Xu}(2021)}]{Xu21}
{Xu}, S. 2021, \apj, 922, 264, \dodoi{10.3847/1538-4357/ac2d8f}

\bibitem[{{Xu} \& {Lazarian}(2018)}]{XLb18}
{Xu}, S., \& {Lazarian}, A. 2018, \apj, 868, 36,
  \dodoi{10.3847/1538-4357/aae840}

\bibitem[{{Xu} \& {Lazarian}(2020)}]{XL20}
---. 2020, \apj, 894, 63, \dodoi{10.3847/1538-4357/ab8465}

\bibitem[{{Xu} \& {Li}(2022)}]{XLi22}
{Xu}, S., \& {Li}, H. 2022, \apjl, 941, L19, \dodoi{10.3847/2041-8213/aca143}

\bibitem[{{Xu} \& {Yan}(2013)}]{XY13}
{Xu}, S., \& {Yan}, H. 2013, \apj, 779, 140,
  \dodoi{10.1088/0004-637X/779/2/140}

\bibitem[{{Xu} {et~al.}(2016){Xu}, {Yan}, \& {Lazarian}}]{Xuc16}
{Xu}, S., {Yan}, H., \& {Lazarian}, A. 2016, \apj, 826, 166,
  \dodoi{10.3847/0004-637X/826/2/166}

\bibitem[{{Yan} \& {Lazarian}(2002)}]{YL02}
{Yan}, H., \& {Lazarian}, A. 2002, Physical Review Letters, 89, B1102+,
  \dodoi{10.1103/PhysRevLett.89.281102}

\bibitem[{{Yan} \& {Lazarian}(2004)}]{YL04}
---. 2004, \apj, 614, 757, \dodoi{10.1086/423733}

\bibitem[{{Zank} {et~al.}(2017){Zank}, {Du}, \& {Hunana}}]{Zank17}
{Zank}, G.~P., {Du}, S., \& {Hunana}, P. 2017, \apj, 842, 114,
  \dodoi{10.3847/1538-4357/aa7685}

\bibitem[{{Zank} {et~al.}(2019){Zank}, {Nakanotani}, \& {Webb}}]{Zank19}
{Zank}, G.~P., {Nakanotani}, M., \& {Webb}, G.~M. 2019, \apj, 887, 116,
  \dodoi{10.3847/1538-4357/ab528c}

\bibitem[{{Zhang} \& {Xu}(2023)}]{ZX23}
{Zhang}, C., \& {Xu}, S. 2023, submitted

\bibitem[{{Zhao} {et~al.}(2020){Zhao}, {Zank}, \& {Burlaga}}]{Zhao20}
{Zhao}, L.~L., {Zank}, G.~P., \& {Burlaga}, L.~F. 2020, \apj, 900, 166,
  \dodoi{10.3847/1538-4357/ababa2}

\bibitem[{{Zirnstein} {et~al.}(2020){Zirnstein}, {Giacalone}, {Kumar},
  {McComas}, {Dayeh}, \& {Heerikhuisen}}]{Zir20}
{Zirnstein}, E.~J., {Giacalone}, J., {Kumar}, R., {et~al.} 2020, \apj, 888, 29,
  \dodoi{10.3847/1538-4357/ab594d}

\bibitem[{{Zirnstein} {et~al.}(2018){Zirnstein}, {Heerikhuisen}, \&
  {Dayeh}}]{Zirn18}
{Zirnstein}, E.~J., {Heerikhuisen}, J., \& {Dayeh}, M.~A. 2018, \apj, 855, 30,
  \dodoi{10.3847/1538-4357/aaaf6d}

\bibitem[{{Zirnstein} {et~al.}(2016){Zirnstein}, {Heerikhuisen}, {Funsten},
  {Livadiotis}, {McComas}, \& {Pogorelov}}]{Zirn16}
{Zirnstein}, E.~J., {Heerikhuisen}, J., {Funsten}, H.~O., {et~al.} 2016, \apjl,
  818, L18, \dodoi{10.3847/2041-8205/818/1/L18}

\bibitem[{{Zirnstein} {et~al.}(2019){Zirnstein}, {McComas}, {Schwadron},
  {Dayeh}, {Heerikhuisen}, \& {Swaczyna}}]{Zim19}
{Zirnstein}, E.~J., {McComas}, D.~J., {Schwadron}, N.~A., {et~al.} 2019, \apj,
  876, 92, \dodoi{10.3847/1538-4357/ab15d6}

\end{thebibliography}

\end{CJK*}
\end{document}